# PV Modules and Their Backsheets -
# A Case Study of a Multi-MW PV Power Station

## Claudia Buerhop-Lutz*, Oleksandr Stroyuk*, Tobias Pickel,
## Thilo Winkler, Jens Hauch, Ian Marius Peters

*High Throughput Methods in Photovoltaics, Forschungszentrum Jülich GmbH, Helmholtz Institute Erlangen-Nürnberg for Renewable Energy (HI ERN), Immerwahrstraße 2, 91058 Erlangen, Germany*

**Authors for correspondence**:

*Dr. Claudia Buerhop-Lutz, High Throughput Methods in Photovoltaics, Forschungszentrum Jülich GmbH, Helmholtz Institute Erlangen-Nürnberg for Renewable Energy (HI ERN), Immerwahrstraße 2, D-91058 Erlangen, Germany; *e-mail*: c.buerhop-lutz@fz-juelich.de

*Dr. Oleksandr Stroyuk, High Throughput Methods in Photovoltaics, Forschungszentrum Jülich GmbH, Helmholtz Institute Erlangen-Nürnberg for Renewable Energy (HI ERN), Immerwahrstraße 2, 91058 Erlangen, Germany; *e-mail*: o.stroyuk@fz-juelich.de

## Abstract

Degradation of backsheets (BS) and encapsulant polymer components of silicon PV modules is recognized as one of the main reasons for losses in PV plant performance and lifetime expectations. Here, we report first insights into the correlation between BS composition of PV-modules and PV power station performance by using a combination of lab- and field-imaging, as well as spectroscopic and electrical characterizations. Using field-suitable near-infrared absorption (NIRA) spectroscopy, the BS structure of 518 PV-modules, 2.5% of the PV-modules in a 5 MWp PV power station, was identified on-site. The variance of the BS composition was found to be heterogeneous across PV-modules of the same power class from the same manufacturer. Polymaide-based BS cause in 10 out of 100 inverters ground impedance values below 400 kΩ, which is a typical threshold for inverters connecting to the grid. For primer-based BS this low value is reached 20 times. We conclude these numbers best as possible from the available monitoring data of inverters associated with BS-types from NIRA. Challenging is the identification of degraded primer-based BSs, since visually they look healthy and undistinguishable to well performing BSs. The present results demonstrate that a deeper understanding of the relationship between bill-of-materials and performance of PV-modules is necessary to avoid/minimize inverter shut-downs, and that it can be achieved by using combinations of selected field and lab characterization methods with monitoring data.

# 1 Introduction

Polymeric components of silicon PV-modules - transparent encapsulants and air-side backsheets (BSs) perform important functions for safe operation of PV power stations. They protect the fragile solar cells from mechanical damage and degradation due to corrosion at ambient operating conditions, and they provide electrical insulation from the environment. A loss in chemical stability, mechanical strength, and electrical insulation properties of BSs and encapsulants results in the deterioration of their protective functions [1-4]. As a result, the long-term stability and durability of PV-modules, and respectively the expected 20-year or more lifespans of PV power stations become compromised.

Various changes of the properties of polymer materials, in particular ethylene vinyl acetate (EVA) encapsulant and BSs of different types were detected in field-aged PV-modules and have been discussed in literature throughout the last years. Variations of BS modifications include discoloration (yellowing, browning), delamination, peeling, macroscopic crack structures, burn marks, corrosion, *etc* [1,2, 4-11].

BSs found in PV-modules can be classified into three groups. The BSs of the first class are composed of a single component, polyamide (PA), while the BSs of the second and third classes are multi-layer BSs. The multilayer BSs comprise a polyethylene terephtalate (PET) core layer. The second class has a symmetric layer structure, that means at the inner layer as well as on the airside layer is a fluorinated polymer. In Contrast, the third BS class has an asymmetric structure, a PET core layer, a single fluorinated layer at the airsides and inner layers of polyolefines, such as polyethylene (PE), polypropylene (PP). The schematic structures of these three BS types, designated as non-fluoropolymer (NF), single-fluoropolymer (SF), and double-fluoropolymer (DF) are shown in Fig. 1. Since the layout of especially multi-layer BSs can vary strongly, we address different configuration as BS-types belonging to a certain BS-class.

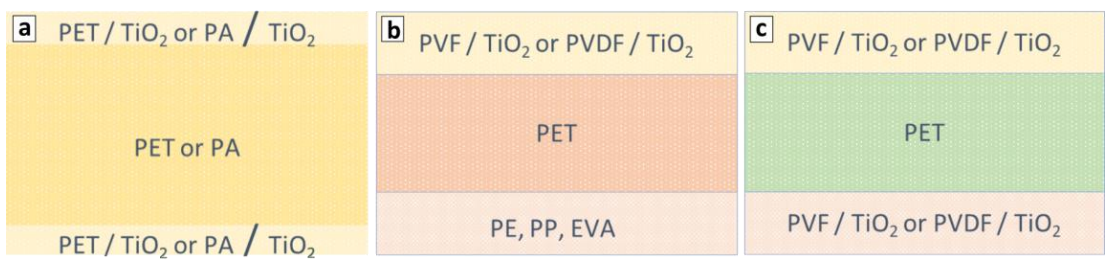

**Figure 1**: Schematic presentation of cross-sections of NF- (a), SF- (b), and DF- (c) BS classes

Gradual recognition of the fact that the degradation behavior of PV-module BSs depends strongly on the BS composition resulted in focused attention on BS failures specific for different polymer materials [1,2]. For example, many reports discuss particular issues of PA-based BS, including "chalking" (titania deposits), massive cracks in the BSs, a loss of mechanical strength and wet leakage resistance as well as potential reasons for degradation of PA-BSs [12-14]. In these studies, most analyses are carried out for single PV-modules in test labs. Typically, the dynamics and mechanisms of the degradation of PV-relevant polymers are studied at defined stress situations in order to rank the PV-modules for long-lasting operation at field conditions and to predict the degradation rate of BS materials at particular conditions [4,11,15-19].

The methods used for characterizing BS materials are mainly combinations of chemical lab analysis with non-destructive BS surface analysis Fourier-transform infrared –FTIR- and Raman spectroscopy [20,21]. At the same time, field studies of polymer-related degradation of PV-module performance are rarely published [1,2,15,22,23]. Such studies are of utmost importance due to the fact that early polymer failures can cause unexpected operational issues beside optical and safety issues (colouring, scratches). Operators claim increasing number of ground faults, inverter alerts, and inverter shut-downs due insulation issues. If the insulation resistance of inverters becomes too low (below 400 kΩ) [24], the inverter does not connect to the grid. There are various reasons for impeding the grid connection, such as damaged PV-modules or DC-wiring, water ingress or damp condensation in junction box, DC isolator, or enclosure. If due to these reasons a low value for the ground impedance is measured, inverters shut down. Not feeding in results in high performance losses and income losses. Thus, degraded and water susceptible BS can cause insulation issues, inverter shut-downs and result in financial losses.

Information about the composition of BS materials in installed PV-modules is scarce and even less is known about the performance and degradation of different BS materials under operating conditions with stress factors like UV-radiation, high temperatures, temperature cycles, humidity, pollution, electrical bias, and material interaction. Deeper insights into how BSs of different layouts degrade in the field, and how this affects PV power station performance is needed. Therefore, BSs must be analysed on-site using field-suitable methods.

The objective of this study is to investigate the field performance of PV-modules depending on different BS types and their impact on PV power station performance and inverter operability. The importance of material-dependent leakage resistance will be presented and discussed with regard to the meaning and the effect on inverter operation as well as its impact on the ground impedance.

For a better understanding of interacting factors and the role of BSs for PV power station performance and operation, we combine six data sets and analyzing methods: 1) lab and field measurement for chemical analysis of the BSs using near-infrared absorption (NIRA) and Raman spectroscopy [25], 2) power measurements via IV curve measurements, 3) wet leakage resistance measurements to assess the insulation state, 4) IR imaging for power-affecting fault detection at operating condition, 5) electroluminescence (EL) imaging for assessing module failures, 6) monitoring data for studying ground impedance of the inverters.

We provide first insights into the correlation between composition / bill of materials (BOM), of PV-modules, with a focus on BS materials, and PV power station performance due to inverter operation in terms of uptime. We especially explore the relevance of externally visible indicators of failures, ageing and degradation of BS and encapsulant as well as changes in performance.

## 2  Approach

The study concentrates on the analysis of one multi-MWp PV power station located in Eastern Germany commissioned in 2012. The analyzed PV capacity is about 5 MWp. More than 20.000 PV-modules from one of the top-ten PV-module manufacturers are installed with various power classes ranging from 225 Wp up to 260 Wp, marked PV-module type A to G. The distribution of the PV-module types is listed in Table 1.

During the inspection of the PV power station in September 2020, many measurements, including aerial thermal imaging, EL imaging, measurements of IV curves on module- and string levels, ground impedance evaluations on inverter level using monitoring data, and NIRA characterization of BSs of single modules and entire strings supported by a visual inspection were carried out on-site.

Table 1: Distribution of the PV-module types and the measured PV-module samples

| MODULE TYPE | A | B | C | D | E | F | G | TOTAL |
|---|---|---|---|---|---|---|---|---|
| INSTALLED PV-MODULES | 250 | 5600 | 100 | 6400 | 6800 | 100 | 1500 | 20750 |
| PV-MODULES WITH ANALYSED BS | 23 | 20 | 22 | 159 | 25 | 15 | 254 | 518 |
| EL-IMAGES | - | 81 | - | - | - | 66 | 321 | 468 |
| IV-CURVES | - | 23 | - | 68 | - | 22 | 161 | 274 |
| IR-IMAGES | 250 | 5600 | 100 | 6400 | 6800 | 100 | 1500 | 20750 |

## 2.1 Field measurements:

Using *aerial IR-imaging* – thermography - a complete inspection of the PV power station was performed in order to identify power relevant failures [26-28], e. g. PV-modules with potential induced degradation (PID), bridged modules, bypassed substrings, *etc*. IR-imaging was performed by a DJI Zenmuse XT2 camera mounted on a DJI M210 drone from an altitude of about 15 m. IR- and IV-measurements were carried out at almost clear-sky conditions, irradiance > 600 W/m², ambient temperatures 20 – 30°C. Further information about the IR-imaging procedure and data processing can be found in [28]. *IV curves* were measured both on module- and on string level using the IV-tracer PVPM 1000CX. *EL images* were taken manually at night with an InGaAS SWIR camera (IRCAM VELOX 327 kN) mounted on top of an extendable pole. Two currents at 20 % and 100 % of the rated short-circuit current $I_{sc}$ of the modules were applied, a detailed description of the measurement procedure can be found in [29].

The values of *ground impedance* (GI) of selected inverters were extracted from monitoring data, giving evidence to when the inverter and was connected to the grid and started feeding-in [24]. Data from roughly 300 inverters with three mpp-trackers were collected. Among them there are inverters with single-variety PV-module types as well as with two or three mixed PV-module types. GI was analyzed if more than ten inverters of identical configuration existed.

Additionally to instrumental measurements *a visual inspection* of the PV-module strings was performed to collect statistics on the installed module power class, population of modules with cracked BSs, disconnected modules.

*NIRA measurements* were performed on 518 single modules from different power classes. The selection of samples was driven by practical measurement aspects, and we selected PV-modules with and without visible anomalies, or anomalies found in IR-images. We collected data from PV-module with all possible BS types found in the PV power station. Using a portable FT-NIR Rocket 2.6 spectrometer (Arcoptix), spectra were collected in reflection mode as discussed in detail in [25].

In total, the recorded data include IR-images of modules with a capacity of 5 MWp, EL-images of modules in 26 strings at high- and low current, 28 string IV-curves, 274 module IV-curves, and 518 BS-analyses, corresponding to 2.5% of the installed PV-modules, as summarized in Table 1. Please note, that in the lab we had samples of all three BS classes, whereas in the field analysis we identified  just BSs of two out of the three introduced classes.

## *2.2 Lab measurements:*

Lab measurements were performed simultaneously with the field measurement campaign and were aimed to identify the BS types met in the field as well as to characterize the insulation state of selected modules. The lab-based part of the study included Raman/NIRA spectroscopic characterization of the structure and composition of BS samples, and wet leakage resistance tests.

*Raman/NIRA characterization*. For spectral IR characterization 35 BS samples of BS cross-sections were cut from discarded modules for Raman/NIRA measurements. This set of samples contained all PV-module types met in the field and, therefore, we assume it to be representative of the BS population on the studied site.

The set of BS cross-sections was inspected with a combination of IR spectro-microscopic approaches. On the first step, all polymer components and pigments were identified by Raman and/or FTIR spectroscopy. Then, multi-spectral Raman maps of BS cross-sections were constructed by confocal Raman microscopy of BS cross-sections [25], visualizing. The maps visualize composition, number, order and thickness of layers and pigments in each particular BS type. Finally, NIRA spectra of each BS type were collected and analyzed. By relating BS structures revealed by Raman mapping with the corresponding NIRA spectra, we could then use exclusively non-destructive NIRA measurements to analyze BSs in the field. We note that all NIRA spectra registered in the field can be assigned to one of the BS types identified by Raman inspection, additionally confirming that the set of 35 cross-sections probed by Raman imaging is representative of the entire studied PV power station.

*Wet leakage resistance*. Wet leakage resistance (LR) was measured by applying a 1000 V bias to PV-modules totally immersed in an aqueous electrolyte bath at 24-25ºC [30]. The leakage current was measured between module surface and a test electrode placed 10 cm from the module into the electrolyte bath. 60-cell PV-modules with polyamide (19 samples) and primer-based BSs (18 samples) were tested and compared to reference modules with 60-cell double-fluoropolymer BSs as discussed below. To obtain kinetic curves of LR losses, 10 PV-modules were kept submersed in the electrolyte for many days, the LR values measured hourly for the first day and then in 6-12 h intervals. In these experiments, PV-modules were intentionally subjected to much harsher conditions than can be met both during the natural exploitation and in the standard wet leakage resistance tests which only require rinsing of the PV-module surface with electrolyte and only for the period of the measurement.

# 3  Results

## 3.1 Lab measurements

### 3.1.1  BS-analysis: Raman and NIRA

*Spectral identification of BS composition.* Typical Raman spectra of BS air-sides show intense bands at 200-750 cm$^{-1}$ characteristic of rutile TiO$_2$ pigment (see spectra in Fig. 2, left part). In particular PA shows characteristic N-H (3300 cm$^{-1}$) and amide C=O (1620 cm$^{-1}$) vibrations, while fluorinated air-side polymers, such as polyvinyl fluoride (PVF) and polyvinylidene fluoride (PVDF) were recognized by characteristic C-F vibrations (1000-1200 cm$^{-1}$) and differences in the C-H vibration series (2700-3000 cm$^{-1}$). Some of the samples showed the presence of a very thin air-side fluorinated polymer layer which could be identified with FTIR spectroscopy revealed a series of C=O- and C-F-related vibrations characteristic for fluorinated polyurethane-polyacrylate copolymers called "primers".

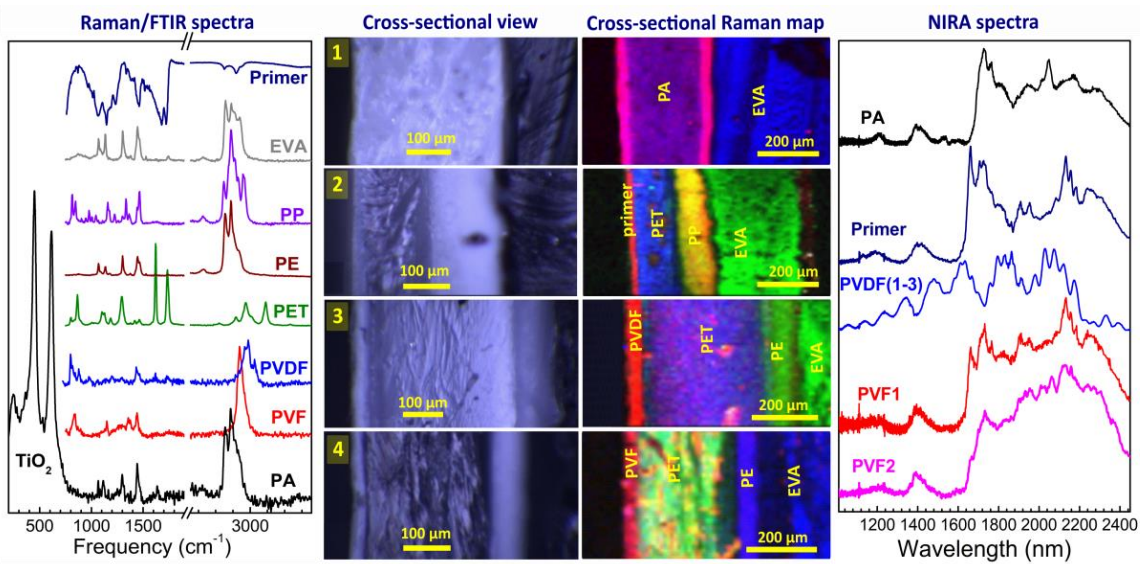

**Figure 2**. Left section - Raman spectra of typical components of BS cross-sections as well as FTIR spectrum of primer air-side (left). *Note*: rutile range is not shown for PVF, PVDF, PP, and PE. Middle section - cross-sectional optical images and multi-spectral Raman maps for different air-side materials. Right section - NIRA spectra of some of tested BSs and EVA encapsulant.

PET intermediate (core) layers were identified by a characteristic C=C/C=O doublet at 1620/1730 cm$^{-1}$. Inner layers of polyolefines (PE or PP) were found in multi-layer BSs, PE in the modules with PVF air-sides and PP in the modules with primer air-sides. The BS cross-sections were cut for Raman analysis along with a layer of EVA encapsulant which can be recognized by a characteristic C=O vibration of acetate fragments at 1730 cm$^{-1}$.

The spectroscopic Raman data were used to build multi-spectral Raman maps of the BS cross-sections [25]. For this, areas of BS cross-sections were scanned pointwise (for example, 100x100 points with a step of 10 µm) taking the Raman spectrum in each point. Then, 2D maps of cross-sections were obtained which resemble optical microphotographs of cross-sections (see middle part in Fig. 2) but allow to distinguish different chemical species. For example, the multi-spectral map of the PA BS shows that the polymer is enriched with rutile at the edges and depleted with TiO$_2$ in the middle section. Mapping of the primer air-side layer in rutile vibrations allows to visualize this layer despite its small thickness (case 2). The Raman mapping of PVDF-based (case 3) and PVF-based (case 4) BS cross-sections allowed BSs varying in the thickness of fluorinated air-side polymer, PET and PE layers to be distinguished.

Table 2 provides a summary of BS identification using the multi-spectral cross-sectional Raman mapping. Totally, we differentiated seven distinct BS types differing by composition and/or structure. Among the fluorinated air-side materials, the primer-based BSs have the thinnest air-side layer and relatively thin intermediate PET component. Other multi-layer BSs are more robust with 30-50 µm air-side layers and much thicker core PET layers (except for PVF/PET/PE-1). In general, the thickness of BSs varies from 300 to slightly above 500 µm.

*Table 2: Overview of layer thickness in µm and structures found on the tested solar field*

| BS class | NF | SF | SF | SF | SF | SF | SF |
|---|---|---|---|---|---|---|---|
| Layer thickness | PA | Primer/ PET/ PP | PVDF/ PET/ PE-1 | PVDF/ PET/ PE-2 | PVDF/ PET/ PE-3 | PVF/ PET/ PE-1 | PVF/ PET/ PE-2 |
| air-side layer (µm) | - | 5-10 | 30 | 30 | 50 | 30 | 40 |
| intermediate layer (µm) | - | 160 | 260 | 330 | 380 | 160 | 260 |
| inner layer (µm) | - | 140 | 120 | 80 | 100 | 110 | 50 |
| Total (µm) | 340 - 390 | 305-310 | 410 | 440 | 530 | 300 | 350 |

For each of the BS types, we collected air-side NIRA spectra which show a number of characteristic features and/or differences for each particular BS type (Fig. 2, right part). In particular, PA-based BSs show a characteristic C-N-H band at 2045 nm.

The PET core of multi-layer BSs can be clearly observed by characteristic =C-H bands at 1660 nm and 2100-2200 nm. The C-F vibrations of fluorinated components do not contribute to NIRA spectra. These components can only be seen by non-characteristic vibration bands of aliphatic -C-H fragments, for example, at 1730-1760 nm.

The PVDF-based BSs show a unique / specific interference pattern in their NIRA spectra. The NIRA spectra of PVF-based BSs are similar to that of primer-based ones, with PET-related features expressed much weaker for PVF1 (PVF/PET/PE-1) because of the differences in the thickness of air-side layer. The signs of PET are even weaker for PVF2 (PVF/PET/PE-2) with the thickest air-side layer of all tested samples.

### 3.1.2 Wet leakage resistance

A reduction in wet LR is observed upon the electrolyte immersion for any kind of PV-module, but the reduction extent varied strongly for different BS types. In particular, PV-modules with the more robust DF BSs showed only about 10% LR reduction after 48 h of immersion (Fig. 3a, curve DF). PV-modules with polyamide-based NF BS revealed typically 40-50% LR reduction after 48 h (curve NF), while the PV-modules with primer-based SF BSs lost more than 80% of the original wet LR (curve SF).

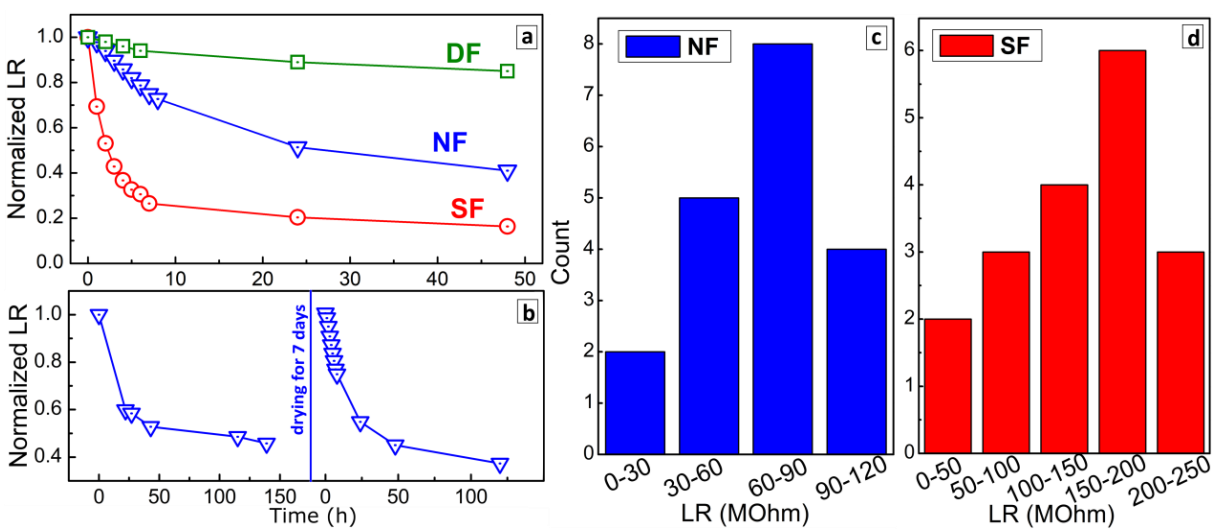

**Figure 3**. (a) Kinetic curves of LR loss during the electrolyte immersion of PV-modules with primer-based SF BS (1), a PA-based NF BS (2), and a DF BS (3). (b) Changes of wet LR for a PV-module NF BS during two cycles of electrolyte immersion interrupted by a 7-day drying pause. (c,d) distributions of wet LRs of PV-modules with NF BSs (c) and SF BSs (d) measured immediately after the electrolyte immersion.

All PV-modules showed negligible variations in J-V curves after the 48-h immersion, indicating that the loss of LR by itself does not result in immediate power losses. Leakage resistance is indicative of chemical changes inside the BS, which can facilitate the degradation of solar cells but is not directly linked an immediate power loss. Perspectively, power loss is possible, little knowledge exits about the timescale.

Wet LR losses were found to be reversible. On the example of a NF BS-based PV-module, we found that LR returns roughly to the original value after a 7-day drying of the module at 24-25ºC and 60% humidity but then drops again as the module is immersed into the electrolyte for the second time (Fig. 3b). This reversible wet LR change is observed for many "immersion-drying" cycles supporting our assumption on a humidity-driven cyclic effect of insulation resistance of PV-modules.

We collected also a set of LR values measured immediately after the electrolyte immersion for a series of PV-modules with NF and SF BSs (Fig. 3c,d). In general, the SF-type modules show higher LR values than NF-based ones, but reveal a much broader LR distribution, from about 40 to 220 MΩ, as compared to NF-type PV-modules (35-95 MΩ). We can take this fact as an additional evidence that SF-type BSs are more susceptible to humidity-related failures than NF-type BSs.

## 3.2 Field measurements

### 3.2.1 BS-analysis: visual inspection, aerial thermography, NIRA

*Distribution of BSs over the strings*. By using air-side NIRA spectroscopy we were able to collect BS data on 518 PV-modules, see Table 1, probing single PV-modules in different strings as well as almost entire strings.

Among the tested ("healthy" looking and suspicious looking) PV-modules, two of the three main BS-classes were found: SF-BS and NF-BS, see Table 3. PA-based NF-BSs and SF-BS are almost equally distributed. Within the SF-BS class seven BS-types differing, e. g. in composition and layer thicknesses, were identified, as shown in Fig. 4a. However, the primer-based BS types dominate the SF-BS class with a share of 63%. The PVDF- and PVF-based SF-BSs constituted minor BS groups on the tested solar field, accounting each for 19%, respectively 15%, of the BS population, we tested. No BSs of DF class were found on the studied PV power station.

The second observation made on the basis of NIRA tests is the variety of possible BS types within power classes. As many as seven BS were found within the same power class (Fig. 4a) Power classes B and D featured five and six different types of BSs.

| MODULE TYPE | A | B | C | D | E | F | G |
|---|---|---|---|---|---|---|---|
| NF-BS: PA | - | - | - | - | - | 100% | 94±2% |
| SF-BS: PVF-BASED | 96±2% | 40±16% | - | 12±1% | - | - | - |
| SF-BS: PVDF-BASED | - | 30±15% | 9±2% | 18±1% | 44±22% | - | 1±1% |
| SF-BS: PRIMER-BASED | 4±2% | 30±15% | 91±2% | 86±1% | 56±22% | - | 4±2% |
| DF-BS | - | - | - | - | - | - | - |

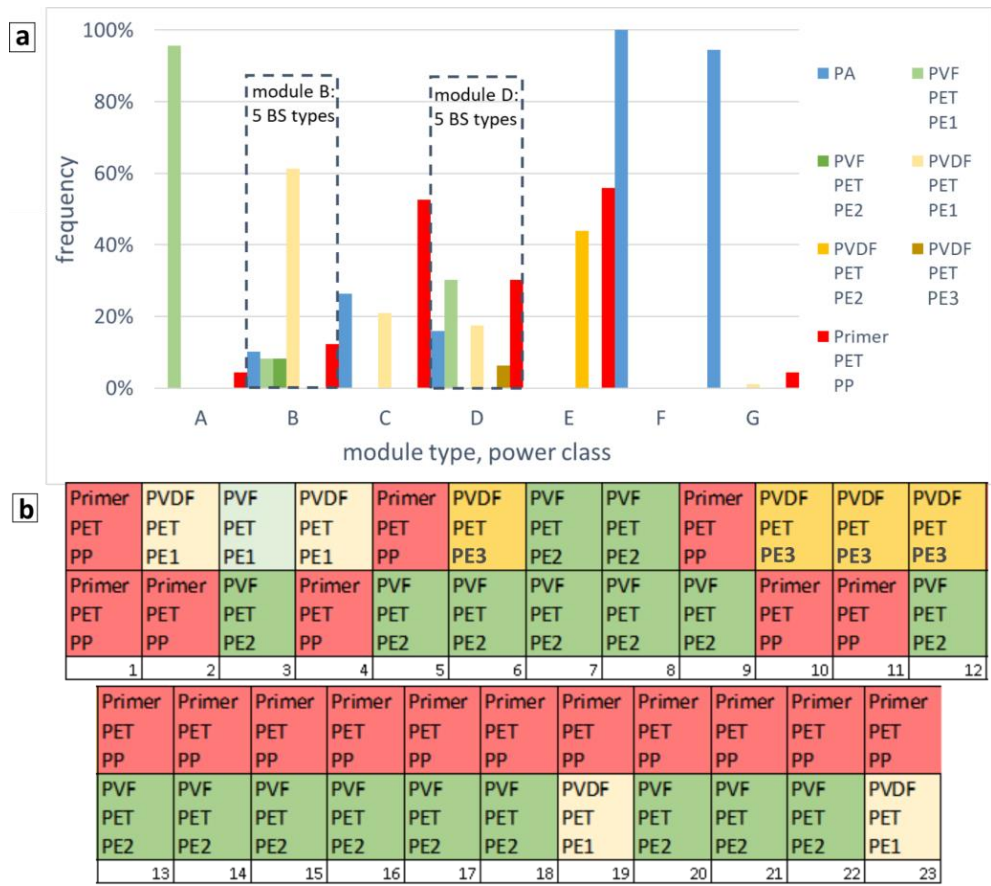

**Figure 4**. (a) Distribution of BS types over the PV-modules of different power classes. (b) An example of two strings (23 modules in each) showing the extreme versatility of BS types for PV-modules type D.

The higher power classes, F and G, were more homogeneous featuring mostly NF-type BSs: PV-module type F 100% PA-based BSs, PV-module type G 94% PA-based BSs and 6% SF-BSs.

For selected PV inverters, we characterized entire PV-module strings and found that a variety of PV-modules of the same power class but with different BS types were united into single strings (see examples in Fig. 4b). Such mixed strings can also potentially be subject to failures because of differences in rates and mechanisms of BS degradations that can cause separate PV-modules to fall out of performance earlier than those with more robust BSs. Seven different BSs is an extreme case, while at least two types of BS were found regularly in the inspected PV-module strings.

*Polyamide BSs - defects and failures*. A visual inspection of PV-modules with PA-based NF BSs revealed numerous cases of extensive cracking of the BS air-sides, approx. 50%-60% of investigated PA PV-modules, and strong chalking (Fig. 5a). Typically, cracks were observed along the metal busbars hidden behind the BS (Fig. 5a). This type of PA cracking was observed throughout the solar plant indicating the same mechanism of PA degradation. Such behavior appears to be of general character for PA-BSs of silicon PV-modules [14]. A cross-sectional study of a cracked BS showed that the cracks to be hundreds of micrometers wide (Fig. 5b) and going through the entire BS till the layer of EVA encapsulant (Fig. 5c). At the same time, a NIRA inspection found no spectral differences between PA-BSs with and without cracks.

An aerial drone-based thermographic inspection of the test site performed simultaneously with NIRA measurements - revealed numerous cases of potential-induced degradation (PID, Fig. 5d). A NIRA study of PID-affected PV-modules revealed that they have predominantly PA-based BSs (32% of PV-modules G are strongly affected by PID).

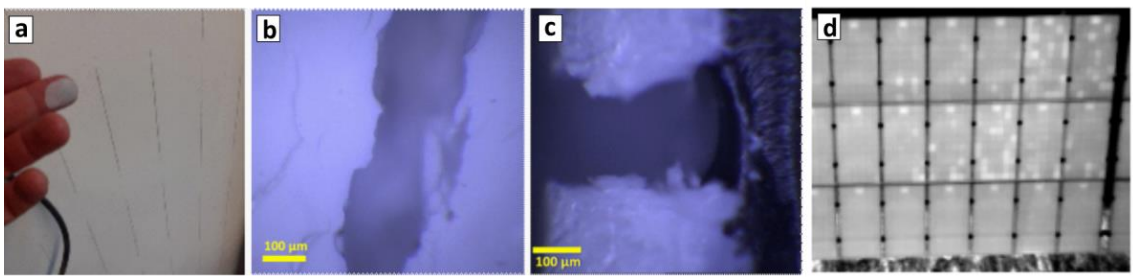

**Figure 5**. (a) Photograph illustrating cracking and "chalking" phenomena typical for PA-BSs. (b,c) microphotographs of a crack in a PA-BS take from the air-side (b) and BS cross-section (c). (d) Aerial thermographic image of a table section with PID-affected PV-modules with PA-BSs.

Almost all tested strings with PV-modules of power classes F and G having PA-BSs showed signs of PID with the number of affected PV-modules close to 25% of all tested PA-based PV-modules, while only 6% of PV-modules with other BS types were affected by PID. At that, the signs of PID were detected for PV-modules with both intact and cracked PA BSs.

*PET-based PV-modules - defects and failures*. Other than PV-modules with PA-BS, PV-modules with SF-type BSs typically show uniform glossy air-sides with no macroscopic cracks on the air-side, but reveal inner cracks when observed using transmitted solar light (Fig. 6a). The inner cracks were assumed to be located either in the PET core layer or in the inner polyolefine layer.

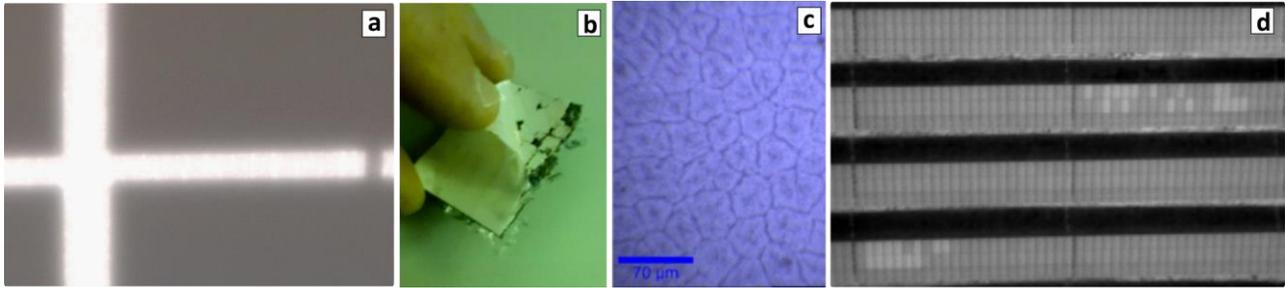

**Figure 6** (a) Photograph of a PV-module with a primer-based SF-type BS taken in transmitted solar light. (b) Photographs of a cut section of a discarded PV-module with a primer-based SF-BS showing signs of drastic degradation of the inner polyolefine layer. (c) Microphotograph of the air-side of a SF-BS showing extensive micro-cracking. (d) Aerial thermographic photograph of a section of solar field showing multiple instances of "bridged" PV-modules.

The latter assumption is also supported by observations made on the samples cut from discarded PV-modules with primer-based BSs intended for cross-sectional Raman studies. Many of these samples showed drastic deterioration of the interior polyolefine layers, which was degraded to a crumbling condition resulting in the delamination of the PET component (Fg. 6b). As these effects were observed for discarded non-functional PV-modules we may conclude this state to be a serious degradation state of the inner layer, observed for other functional PV-modules on the initial stage as the internal cracking (Fig. 6a).

The air-side of the primer-based SF-BSs typically shows a micro-cracking pattern (Fig. 6c), that is, a surface network of microcracks (several micrometers wide) making the internal layers potentially accessible to humidity and atmospheric gases.

The aerial inspection of tables composed of PV-modules with SF-type BSs showed that a considerable amount of PV-modules were permanently removed from the string/bypassed ("bridged", Fig. 6d), most probably, to avoid inverter shut-downs due to insulation issues in these PV-modules. The bridging phenomenon was mostly noticed for PV-modules with the primer-based SF-BSs, that is, those having the thinnest and microcracked fluorinated polymer layer on the air-side.

## 3.2.2 Monitoring data analysis

Evaluating monitored GI values gives evidence for up-coming insulation issues in the PV

power station. The data in Fig. 7a show the minimum daily GI for inverters with different PV-module types: inverters with PV-modules G predominantly with PA-based BS (Fig. 7b) and inverters with PV-modules D and E mostly with SF-based BS (Fig. 7c).

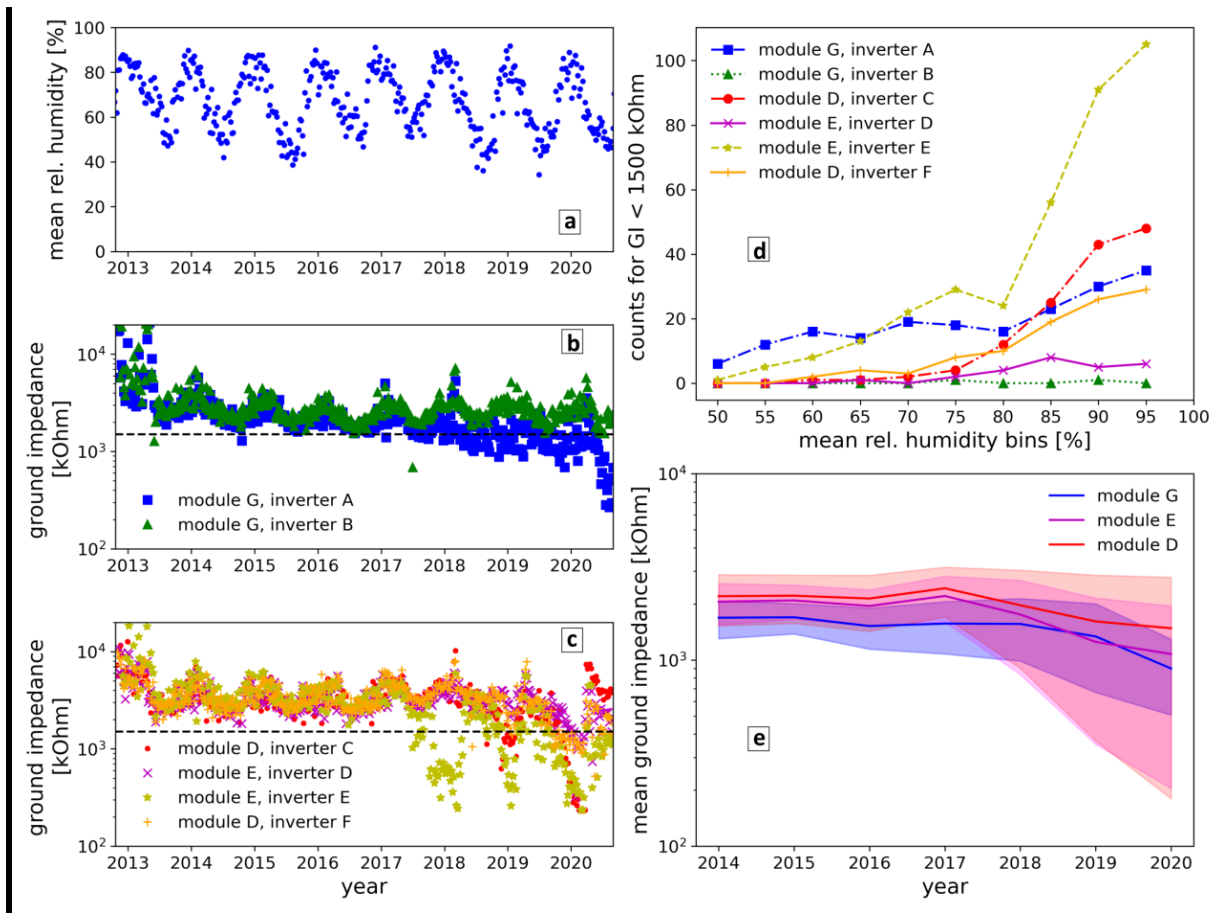

**Figure 7**: Time-series of minimum ground impedance for various PV-module types and differing BSs, a) mean daily relative humidity, b) inverters of PV-modules G with most likely a PA-BS, c) inverters of PV-modules B and D with most-likely SF-BS (PV-modules D, inverter C corresponds to the inverter shown in Fig. 8, PV-modules D, inverter F corresponds to the inverter shown in Fig. 5b), d) influence of mean daily relative humidity on ground impedance for 2013 to 2020, and e) mean ground impedance for the populations of PV-modules D, E, and G.

In the first years of operation the minimum daily GI was constantly above 1500 kΩ, independent on the weather conditions, irrespective of the humidity [31]. After several years, GI starts to react to humidity, with the exact reaction depending on the BS-type. The GI of PV-modules with NF-BS varies between remaining constant or decreasing slowly but constantly. Here, the drop of GI below 1500 kΩ seems to be dependent of humidity after the onset in the fifth year of operation. The peak at high humidity for inverter A may be due to the fact that here at least four modules F have a primer-based BS (known from the NIRA-analysis). In contrast, inverters with PV-modules with mainly SF-BS deviate in performance, see Fig. 7c, the GI for inverters with module types D and E varies strongly between constantly high (comparable to those with PA-based BS, Fig. 7b) or spontaneously decreasing strongly, especially during the more humid winter seasons. The drop of GI depends strongly on the humidity, as Fig. 7d shows. At days of high humidity, GI drops often below 1500 kΩ but recovers for dry day of low humidity.

The distribution for the minimum ground impedance for the populations of module types D, E, and G in shown in Fig. 7e. At that, 20% of the inverters with almost exclusively NF-BS PV-modules reach values lower than 1000 kΩ, only 11% fall under 400 kΩ, although BS cracks were seen in at least 60% of these PV-modules. The GI drop is moderate.

Considerable amounts of inverters with PV-modules D (44.5%) and E (47.8%) reached as low values as 1000 kΩ. GI values lower than 400 kΩ were registered for 20% of the inverter with PV-modules D or E, signalling on severe insulation issues. As data in Table 4 point out, severe GI drop is observed for module types with mainly identified SF-BSs.

As a result, solving the insulation issues by bridging / removing PV-modules with low $R_{iso}$-values from the string, becomes a common practice. This increase GI, at least in a short-term (see Fig. 7c, red dots).

*Table 4: Summary of ground impedance analysis (inverter is counted if the threshold is reached at least 10 days throughout the operation time of 2434 days)*

| Inverter with | expected BS-class (BS-type) | # of inverters | # of inverters with GI < 1500 kΩ | # of inverters with GI < 400 kΩ |
|---|---|---|---|---|
| PV-module type F+G | NF | 18 | 12 (66%) | 2 (11%) |
| PV-module type D | SF | 23 | 5 (30%) | - |
| PV-module type E | SF | 92 | 55 (60%) | 18 (20%) |
| PV-module type B+D | SF | 138 | 90 (56%) | 33 (20%) |

Even though BS-types of all PV-modules of the intensively studied inverters are not known, GI of PA-based BSs shows no tendency to fall frequently below 1500 kΩ, see Fig. 8**.** An explanation can be that there are more than four modules with primer-based BS combined with modules with PA-BS, and / or additional issues, e. g. cabling, junction boxes.

For inverter with PV-modules with mixtures of SF-BS-types, e. g. primer-based BS with PVF-based and / or PVDF-based BS, we see indications the GI drop increases with the number of modules with primer-based BSs exceeding eight of such modules per inverter. The worst case observed, is an inverter with 42 modules with primer-based BS. Here the GI was lower than 1500 kΩ for 20% of the total operation days. This GI drop indicates existing or upcoming insulation issues which can lead to future inverter issues as soon as lower values are reached. Most-likely inverters with PV-modules with SF-BS will lead to more alerts due to ground faults than PV-modules with PA-BS do. The collected and analyzed GI data provide valuable information on the BS material and enable first interferences on material-specific properties and performance.

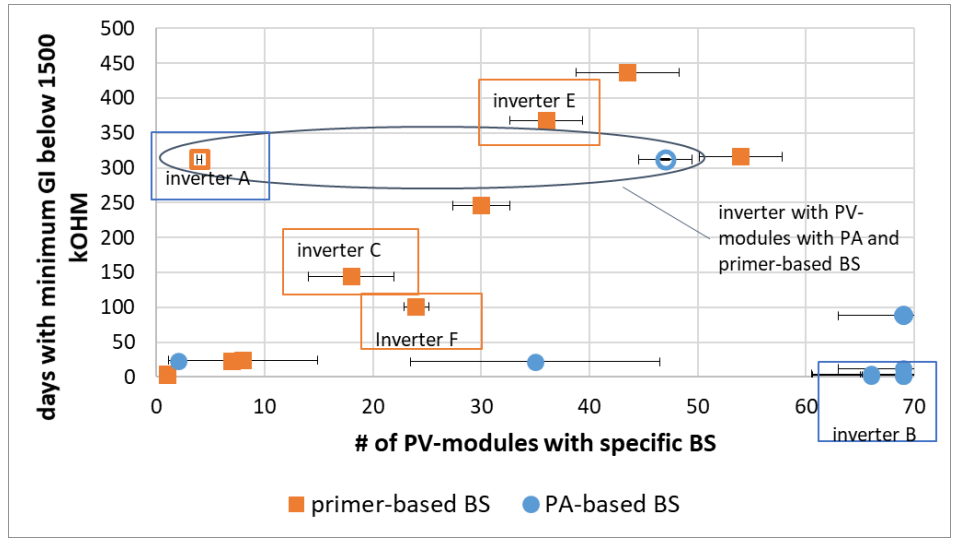

**Figure 8**: Evaluation of GI for inverters with measured and known BS for more than ten PV-modules, error bars visualize the uncertainty of the number of PV-modules due to the varying number of analyzed PV-modules (between 4 and 51 PV-modules per inverter), 2434 days of total operation time.

### 3.2.3 IV-measurements

Due to insulation issues PV-modules have been excluded from the electrical circuit by the operators, in other words these PV-modules are "bridged". An inspection of J-V curves of selected bridged PV-modules showed them to be fully operational. As the measurements were taken on dry and warm late summer season we assumed that the insulation issues resulting in the bridging of the PV-modules should be observed on humid seasons, such as winter or spring, as well as on day times with extensive dew deposition, and should be reversible and not observed during dry and warm times. An additional NIRA inspection of the BSs of the "bridged" PV-modules showed no particular differences between unaffected and "bridged" PV-modules, additionally indicating on a transient character of the failures resulting in the bridging.

Figure 9 illustrates the results of the on-site IV-measurements of PV-modules and strings including bridged PV-modules for two exemplary strings. In IR-images, taken before the IV measurements, the bridged PV-modules, operating at open circuit have a higher temperature than the surrounding, electrically connected PV-modules, and are clearly visible (Fig. 9a).

However, the measured power lies within the range of the normally operating PV-modules (Fig. 9b). When all PV-modules in the tested strings are considered in the string power measurement, the power gain corresponds to the number of bridged PV-modules (Fig. 9c).

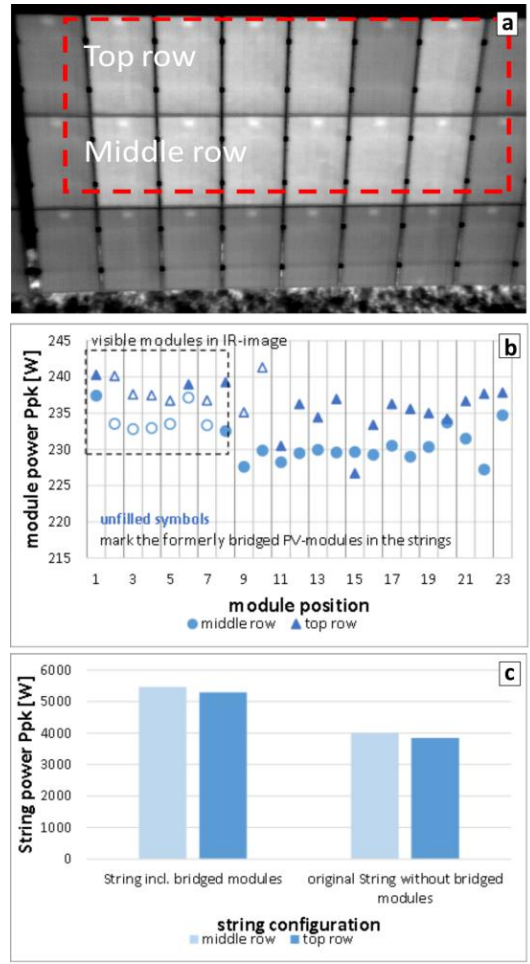

**Figure 9**: IV-measurements of bridged and un-bridged PV-modules in a PV-array with PV-modules D with a mostly primer-based BS, a) IR-image showing bridged PV-modules at an increased temperature, b) measured $P_{pk}$ of bridged and un-bridged PV-modules, and c) comparison of measured string power for the original string (without bridged PV-modules) and the string including the originally bridged -PV-modules (top row: in total seven bridged PV-modules, middle row: in total six bridged PV-modules)

This observation indicates that the insulation problems resulting in the PV-module bridging can only be observed at relevant conditions, for example, at an elevated humidity. We found 137 bridged modules affecting the energy yield of 16 strings. Between one PV-module bridged in one string and the entire string disconnected from the inverter, everything occurs. In line with this expectation, we do observe severe yield losses in the yearly monitoring data when the bridged PV-modules are accounted for along with unbridged ones.

# 4 Discussion

Three BS classes, DF-BS, SF-BS, and NF-BS, were introduced. Examples of two BS classes, SF-BS and NF-BS, were found in the inspected PV power station. Within the sample, 49% of all BSs are NF-BSs made of PA, 51% were various SF-BSs. 60% of the SF-BS from the sample have a primer-based BS. According to these numbers, PA-BSs are overrepresented because PA is only found in PV-modules of type F and G (10% of the population). This enables a first overview of BS-specific findings and their impact on the field performance, as summarized in Table 5, including DF-BS due to lab measurements.

PV-modules with PA-BS show clearly visible signs of degradation as severe cracking and chalking of the BSs occurs. Measuring the IV-curve reveals no significant power loss, though. PID was detected frequently in this module type, see Figure 5, and we find a connection with susceptibility to water ingress, see Figure 3.

*Table 5: Summary of findings for different BS-classes (monitoring data analysis: yield Y and voltage V)*

|       | *Visuals* | *NIRA* | *LR* | *Pmpp* | *Field data* | |
|-------|-----------|--------|------|--------|--------------|---|
|       |           |        |      |        | *IR-imaging* | *Historical monitoring data (Y, V)* |
| *DF-BS* | - | PVF-PET-PVF | High-medium | Unaffected | - | Constant [32] |
| *SF-BS* | micro-cracks | Primer-PET-PP | Medium-low | Unaffected | Bridged modules PV- | Drastic and spontaneous drop |
| *NF-BS* | cracks | PA, PET | Medium-low | Unaffected | PID | Slight decrease |

Wet leakage tests reveal severe loss of leakage resistance, as well as its recovery during subsequent drying. For field operation, BS susceptibility to water is of importance for the inverter operation. After several years of operation the insulation resistance of PV-modules with PA-based BSs starts to decrease slightly but constantly. As a result, ground impedances of inverters go down continuously. In case of modules with single-variety PA-BSs (only one single PS type detected yet), the GI drop seems to be moderate, in case of PV-modules with multi-variety BSs (different BS types found for one module type) we found indications for significant GI-drops. Monitoring data of inverters show this behavior; yet we have no complete list of all BS types, so we conclude this best as possible from the data we have. A complete NIRA scan of all PV-modules would help. However, there are a couple of statements we can already make.

While visual signatures of failures are known for modules with PA-BS, the appearance of modules with degraded SF-BSs is a less obvious sign of their condition. PV-modules with SF-BSs may appear intact and unharmed although the degradation process has already started. On a macroscopic scale, they can easily be mistaken for healthy BSs. Closer inspection reveals degradation patterns, e. g. micro-cracks of the inner layers, at the surface, corrosion of metal components, delamination. These defects enable the ingress as well as the exit of water into and from the material. The strong loss of wet leakage resistance upon immersion confirms this behavior as does the fast recovery after drying. Under operation, we suspect that the high susceptibility to water leads to insulation issues during seasons of high humidity, and may cause ground faults and inverter issues. A further indication of this assumed process is that in summer and on dry days insulation issues disappear [32,33].

In general, the number of faults increases with time, even if weak PV-modules causing the insulation issues are bridged or removed from the string. Others with the same primer-based BS will follow. Because "good" and "bad" BS, more robust and already degraded BS cannot be distinguished by naked-eye, field-suitable characterization techniques, as NIRA, need to be established, to collect more data and improve the quality control. Data of the dynamics of the leakage resistance of other SF-BS types, e. g. PVF-types, PVDF-types besides the primer-based SF-BS, need to be included in the discussion.

Furthermore, there is no direct impact of poor leakage resistance on the module power but indirectly when modules are bridged (which shortens the string), inverters do not connect to the grid or connect later during the day for feeding-in. Income losses will pile up.

# 5  Conclusions

Using the field suitable NIRA-method the BS material of 518 PV-modules, 2.5% of the PV-modules in a 5 MWp PV power station, were identified on-site. The variance of the BS-type with respect to the module type is unexpectedly heterogeneous. The external appearance is not indicative of the degradation state of the BS. Up-coming insulation issues are not captured by frequently used methods, as thermography, EI-imaging, IV-measurements.  The type of BS was found to impact the inverter operation differently and on a different time-scale, slow and steadily for modules with PA-based BS and spontaneously and drastically for primer-based SF class BSs. We conclude this behaviour best as possible from the monitoring data of inverters we have. A complete NIRA scan of all PV-modules would help.

PV power stations with BS issues we identified are mainly commissioned within the boom years 2010 – 2015. Here, 40% of the PV capacity registered in Germany 2020 was installed.

According to our observations, this 20 GW PV capacity has an unknown risk to fail. A significant fraction may not fulfill its expected lifetime due to its backsheet material.

A deeper understanding of the relationships between the composition and structure of the PV-module BSs can help to use early knowledge for countermeasures avoiding / minimizing inverter shut-downs. The existing knowledge of the degradation processes of polymers at ambient operating conditions is scarce and needs to be developed by massive field measurements using combinations of robust field approaches as illustrated here. Accumulation of field-relevant information on the PV-module performance and degradation will allow to achieve a better and more focused selection and optimization of the BS and encapsulant materials for long-lasting PV-modules.

*Acknowledgements*. This work is supported by the German Federal Ministry for Economic Affairs and Energy (project "COSIMA", FKZ: 0324291A); and Zentrales Innovationsprogramm Mittelstand (ZIM, project "PolymerCHECK", No. 16KN083038). We thank Allianz Risk Consulting GmbH for the access to the PV power stations.

*Declaration of conflicts of interests*. The authors declare no conflicts of interests.